\documentclass[pra,showpacs,showkeys,amsfonts,amsmath,twocolumn]{revtex4}
\usepackage{bm}
\usepackage{graphicx}
\RequirePackage{mathptm}

\numberwithin{equation}{section}

\newcommand{\sa}[2]{\sigma_{#1}^{#2}}

\newcommand{\ro}{\rho}
\newcommand{\ras}{\rho_{\mr{as}}}

\newcommand{\g}{\gamma}
\newcommand{\ga}[1]{\gamma_{#1}}
\newcommand{\Ga}[1]{\Gamma_{#1}}
\newcommand{\Om}[1]{\Omega_{#1}}
\newcommand{\re}{\mathrm{Re}\,}
\newcommand{\I}{\openone}
\newcommand{\conj}[1]{\overline{#1}}
\newcommand{\ket}[1]{|{#1}\rangle}
\newcommand{\bra}[1]{\langle {#1} |}

\newcommand{\C}{\mathbb C}

\newcommand{\tr}{\mathrm{tr}\,}

\newcommand{\mr}[1]{\mathrm{#1}}

\begin{document}
\title{Vacuum - induced stationary entanglement in radiatively coupled three - level atoms }
\author{{\L}ukasz Derkacz}
\affiliation{Institute of Theoretical Physics\\ University of
Wroc{\l}aw\\
Plac Maxa Borna 9, 50-204 Wroc{\l}aw, Poland}
\author{Lech Jak{\'o}bczyk\footnote{
E-mail addres: ljak@ift.uni.wroc.pl}} \affiliation{Institute of
Theoretical Physics\\ University of
Wroc{\l}aw\\
Plac Maxa Borna 9, 50-204 Wroc{\l}aw, Poland}
\begin{abstract}
We consider a pair of three - level atoms interacting with a common
vacuum and analyze the process of entanglement production due to
spontaneous emission. We show that in the case of closely separated
atoms, collective damping can generate robust entanglement of the
asymptotic states.\\[4mm]
SHORT TITLE: Stationary entanglement between three - level atoms
\end{abstract}
  \maketitle
\section{Introduction}
The important problem of evolution of entanglement in realistic
quantum systems interacting with their environments was mainly
discussed in the case of two two - level systems (qubits). In that
case, the interesting idea that dissipation can create rather then
destroy entanglement was studied in details. In particular, in the
case of two - level atoms, the possible production of robust or
transient entanglement induced by the process of spontaneous
emission was shown \cite{Ja, FT,TF, JaJa}.
\par
Much more complex and interesting is the process of creation of
entanglement involving multilevel atoms. In such a case, quantum
interference between different radiative transitions can influence
the dynamics of the system. For a pair of largely separated three -
level atoms the role of such interference in the process of
degradation of entanglement was studied in Ref. \cite{DeJa}. When
the interatomic distance is comparable to the wavelength of the
emitted radiation, the coupling between the atoms via common vacuum
gives rise to the collective effects such as collective damping and
dipole - dipole interaction. Such effects are well known \cite{Ag},
particularly in the case of two - level atoms. In the system of
three - level atoms having closely lying excited states, radiative
coupling can produce a new interference effect in the spontaneous
emission. This effect manifests by the cross coupling between
radiation transitions with orthogonal dipole moments \cite{AP}. All
such collective properties of the system influence the entanglement
between three - level atoms.
\par
In the paper, we study the entanglement production between three -
level atoms due to the collective damping (a detailed analysis of
the entanglement evolution in the presence of all collective effects
will be presented elsewhere). In that case, the analysis is involved
since there is no simple necessary and sufficient condition of
entanglement for a pair of $d$ - level systems with $d\geq 3$.
Peres-Horodecki separability criterion \cite{P,3H} only shows that
states which are not positive after partial transposition (NPPT
states) are entangled. But there can exist entangled states which
are positive after this operation \cite{H} (bound entangled PPT
states). The problem of existence of bound entangled (i.e.
non-distillable \cite{HHH}) states can be analyzed in terms of the
rank of density matrix of bipartite system and the ranks of its
partial traces. If a state is separable or bound entangled, then its
rank must be larger then the ranks of partial traces \cite{HSTT}. In
the paper we focus on the possibility of creation of NPPT "free"
entangled states, so we not discuss these problems. To detect and
quantify entanglement we use negativity of partial transposition of
density matrix.
\par
As we show, in the limit of small separation between the atoms, the
process of the photon exchange between the atoms produces such
correlations that the dynamics is not ergodic and there are
nontrivial asymptotic stationary states. We compute the explicit
form of the asymptotic state for any initial state and show that
some of the asymptotic states are NPPT states, even if the initial
states were PPT states. This effect occurs for example for a large
class of diagonal i.e. separable initial states. We give also the
example of bound entangled PPT state which evolves into NPPT (i.e.
distillable) asymptotic state.
\section{Dynamical evolution of two three-level atoms}
Consider two identical three - level atoms ( $A$ and $B$) in the $V$
configuration. The atoms have two near - degenerate excited states
$\ket{1_{\mu}},\; \ket{2_{\mu}}$ ($\mu =A,B$) and ground states
$\ket{3_{\mu}}$. Assume that the atoms interact with the common
vacuum and that transition dipole moments of atom $A$ are parallel
to the transition dipole moments of atom $B$. Due to this
interaction, the process of spontaneous emission from two excited
levels to the ground state  take place in each individual atom but a
direct transition between excited levels is not possible. Moreover,
the coupling between two atoms can be produced by the exchange of
the photons. The evolution of atomic system can be described by the
following master equation \cite{Ag}
\begin{equation}
\frac{d\ro}{dt}=(L^{A}+L^{B}+L^{AB})\ro\label{me}
\end{equation}
where for $\mu=A,B$ we have
\begin{equation}
\begin{split}
L^{\mu}\ro=&\ga{13}\,\left(\,2\sa{31}{\mu}\ro\sa{13}{\mu}-\sa{13}{\mu}
\sa{31}{\mu}\ro-\ro\sa{13}{\mu}\sa{31}{\mu}\right)
+\\[2mm]
&\ga{23}\,\left(\,2\sa{32}{\mu}\ro\sa{23}{\mu}-\sa{23}{\mu}
\sa{32}{\mu}\ro-\ro\sa{23}{\mu}\sa{32}{\mu}\,\right)
\label{genA}
\end{split}
\end{equation}
and
\begin{equation}
\begin{split}
L^{AB}\ro=&\hspace*{3mm}\Ga{13}\,(\,
2\sa{31}{A}\ro\sa{13}{B}-\sa{13}{B}\sa{31}{A}\ro-\ro\sa{13}{B}\sa{31}{A}\\[2mm]
&+2\sa{31}{B}\ro\sa{13}{A}-\sa{13}{A}\sa{31}{B}\ro-\ro\sa{13}{A}\sa{31}{B}\,)\\[2mm]
&+i\,\Om{13}\,\left[\,\sa{13}{A}\sa{31}{B}+\sa{13}{B}\sa{31}{A},\ro\,\right]\\[2mm]
&+\Ga{23}\,(\,
2\sa{32}{A}\ro\sa{23}{B}-\sa{23}{B}\sa{32}{A}\ro-\ro\sa{23}{B}\sa{32}{A}\\[2mm]
&+2\sa{32}{B}\ro\sa{23}{A}-\sa{23}{A}\sa{32}{B}\ro-\ro\sa{23}{A}\sa{32}{B}\,)\\[2mm]
&+i\,\Om{23}\,\left[\,\sa{23}{A}\sa{32}{B}+\sa{23}{B}\sa{32}{A},\ro\,\right]\label{genAB}
\end{split}
\end{equation}
In the equations (\ref{genA}) and (\ref{genAB}), $\sa{jk}{\mu}$ is
the transition operator from $\ket{k_{\mu}}$ to $\ket{j_{\mu}}$ (
$\mu=A,B$ ) and the coefficient $\ga{j3}$ represents the single atom
spontaneous - decay rate from the state $\ket{j}$ \\( $j=1,2$ ) to
the state $\ket{3}$. The coefficients $\Ga{j3}$ and $\Om{j3}$ are
related to the coupling between two atoms and are the collective
damping  and the dipole - dipole interaction potential,
respectively. As was shown in Ref. \cite{AP}, in such atomic system
there is also possible the radiative process in which atom $A$ in
the excited state $\ket{1_{A}}$ loses its excitation which in turn
excites atom $B$ to the state $\ket{2_{B}}$. This cross coupling
between two atoms is sensitive to the orientation of the transition
dipole moments of atoms, and in the present paper we study the model
in which that coupling is absent. We also assume that the
spontaneous - decay rates satisfy
\begin{equation}
\ga{12}\approx \ga{13}=\g \label{gamma}
\end{equation}
The remaining coefficients in equation (\ref{genAB}) can be written
as
\begin{equation}
\begin{split}
&\Ga{j3}=\g\; G_{j3}(R)\\
&\Om{j3}=\g\; F_{j3}(R)\\
\end{split}
\end{equation}
where $j=1,2$ and $R$ is the distance between atoms. A detailed form
of the functions $G_{j3}(R),\; F_{j3}(R)$ depends on the geometry of
the system \cite{AP}, but in general, for $R\to\infty$
$$
 G_{j3}(R),\,
F_{j3}(R)\to 0
$$
and for $R\to 0$
$$
G_{j3}(R)\to 1
$$
whereas the functions $F_{j3}(R)$ diverge.
\par
The time evolution of the initial state $\ro$ of the atomic system
is given by the semi - group $ \{ T_{t} \}_{t\geq 0} $ of completely
- positive linear mappings acting on density matrices \cite{A},
generated by $L^{A}+L^{B}+L^{AB}$. The properties of this semi -
group crucially depend on the distance $R$ between two atoms. As can
be shown by a direct calculations, when the distance is large
(compared to radiation wavelenght), the semi - group
$\{T_{t}\}_{t\geq 0}$ is uniquely relaxing with the asymptotic state
$\ket{3_{A}}\otimes \ket{3_{B}}$. On the other hand, when $R$ is
small, $\Ga{13},\; \Ga{23}\to \g$ and $\Om{13},\; \Om{23}$ are
large, so we can use the approximation
\begin{equation}
\Ga{13}=\Ga{23}=\g\quad\text{and}\quad
\Om{13}=\Om{23}=\Omega\label{approx}
\end{equation}
In that case,  the semi - group is not uniquely relaxing and
asymptotic stationary states are nontrivial and depend on initial
conditions.
\par
We do not discuss details of the time evolution of the system, but
we focus on the analysis of the asymptotic behaviour of the dynamics
of   atoms with small separation, when the conditions (\ref{gamma})
and (\ref{approx}) are satisfied. The master equation (\ref{me}) can
be used to obtain differential equations for the matrix elements of
any state $\ro$. We consider matrix elements  of $\ro$ with respect
to the basis of $\C^{3}\otimes \C^{3}$ given by vectors
\begin{equation}
\ket{j_{A}}\otimes \ket{k_{B}},\quad j,k=1,2,3\label{basis}
\end{equation}
taken in the lexicographic order. The equations for $\ro_{lm},\;
l,m=1,\ldots, 9$ form a system of linear differential equations
which can be solved by elementary methods. Using these solutions,
after a long calculations, we obtain the explicit form of the
asymptotic state $\ras$ for any initial state $\ro$ with the matrix
elements $\ro_{lm}$:
\begin{equation}
\ras=\begin{pmatrix} 0&0&\hspace*{2mm}0&0&0&\hspace*{2mm}0&\hspace*{2mm}0&\hspace*{2mm}0&\hspace*{2mm}0\\
0&0&\hspace*{2mm}0&0&0&\hspace*{2mm}0&\hspace*{2mm}0&\hspace*{2mm}0&\hspace*{2mm}0\\
0&0&\hspace*{2mm}x&0&0&\hspace*{2mm}z&-x&-z&\hspace*{2mm}w\\
0&0&\hspace*{2mm}0&0&0&\hspace*{2mm}0&\hspace*{2mm}0&\hspace*{2mm}0&\hspace*{2mm}0\\
0&0&\hspace*{2mm}0&0&0&\hspace*{2mm}0&\hspace*{2mm}0&\hspace*{2mm}0&\hspace*{2mm}0\\
0&0&\hspace*{2mm}\conj{z}&0&0&\hspace*{2mm}y&-\conj{z}&-y&\hspace*{2mm}v\\
0&0&-x&0&0&-z&\hspace*{2mm}x&\hspace*{2mm}z&-w\\
0&0&-\conj{z}&0&0&-y&\hspace*{2mm}\conj{z}&\hspace*{2mm}y&-v\\
0&0&\hspace*{2mm}\conj{w}&0&0&\hspace{2mm}\conj{v}&-\conj{w}&-\conj{v}&\hspace*{2mm}t
\end{pmatrix}\label{roas}
\end{equation}
where
\begin{equation}
\begin{split}
&x=\frac{1}{8}\,(\ro_{22}+2\ro_{33}+\ro_{44}+2\ro_{77}-2\,\re \ro_{24}-4\,\re \ro_{37})\\
&z=\frac{1}{4}\,(\ro_{36}-\ro_{38}-\ro_{76}+\ro_{78})\\
&w=\frac{1}{4}\,(\ro_{26}+\ro_{28}+2\ro_{39}-\ro_{46}-\ro_{48}-2\ro_{79})\\
&y=\frac{1}{8}\,(\ro_{22}+\ro_{44}+2\ro_{66}+2\ro_{88}-2\,\re \ro_{24}-4\, \re \ro_{68})\\
&v=\frac{1}{4}\,(-\ro_{23}-\ro_{27}+\ro_{43}+\ro_{47}+2\ro_{69}-2\ro_{89})
\end{split}\label{roasdetails}
\end{equation}
and
$$
t=1-2x-2y
$$
To get some insight into the process of creation of non - trivial
asymptotic state $\ras$, it may be useful to consider the  basis of
collective states in $\C^{9}$, given by the doubly excited states
$$
\ket{e_{1}}=\ket{1_{A}}\otimes\ket{1_{B}},\quad
\ket{e_{2}}=\ket{2_{A}}\otimes\ket{2_{B}}
$$
the ground state
$$
\ket{g}=\ket{3_{A}}\otimes\ket{3_{B}}
$$
and generalized symmetric and antisymmetric Dicke states (see e.g.
\cite{barg})
\begin{equation}
\begin{split}
&\ket{s_{kl}}=\frac{1}{\sqrt{2}}\left[\,\ket{k_{A}}\otimes\ket{l_{B}}
+\ket{l_{A}}\otimes\ket{k_{B}}\,\right]\\
&\ket{a_{kl}}=\frac{1}{\sqrt{2}}\left[\,\ket{k_{A}}\otimes\ket{l_{B}}
-\ket{l_{A}}\otimes\ket{k_{B}}\,\right]
\end{split}
\label{dicke}
\end{equation}
where $k,l=1,2,3\,; k<l$. The states (\ref{dicke}) are entangled,
but in contrast to the case of two - level atoms, they are not
maximally entangled. One can also check that the doubly excited
states $\ket{e_{1}},\; \ket{e_{2}}$ and the symmetric Dicke states
$\ket{s_{kl}}$ decay to the ground state $\ket{g}$, whereas
antisymmetric states $\ket{a_{13}}$ and $\ket{a_{23}}$ decouple from
the environment and therefore are stable. Moreover, the state
$\ket{a_{12}}$ is not stable, but it is asymptotically non -
trivial. Notice that the collective states can be used to the direct
characterization of the asymptotic behaviour of the system. In
particular, the parameters $x$ and $y$ in (\ref{roas}) are given by
the populations in the antisymmetric states $\ket{a_{13}},\,
\ket{a_{23}}$ and $\ket{a_{12}}$:
\begin{equation*}
\begin{split}
&x=\frac{1}{4}\left(\,\bra{a_{12}}\ro\ket{a_{12}}+2\,\bra{a_{13}}\ro\ket{a_{13}}\,\right)\\
&y=\frac{1}{4}\left(\,\bra{a_{12}}\ro\ket{a_{12}}+2\,\bra{a_{23}}\ro\ket{a_{23}}\,\right)
\end{split}
\end{equation*}
The remaining parameters can be computed in terms of the coherences
between the collective states. Since the populations
$\bra{a_{13}}\ro\ket{a_{13}}$ and $\bra{a_{23}}\ro\ket{a_{23}}$ are
stationary, the states which have the property of trapping the
initial populations in $\ket{a_{13}}$ or $\ket{a_{23}}$, create the
non - trivial asymptotic state $\ras$ with the stationary
entanglement. On the other hand, the population in the state
$\ket{a_{12}}$ is not stable, but it can be transformed into
$\bra{a_{13}}\ro\ket{a_{13}}$ and $\bra{a_{23}}\ro\ket{a_{23}}$ in
such a way that the values of the parameters $x$ and $y$ are fixed.
The explicit examples of such behaviour of the system will be
discussed in next section.
\section{Generation of NPPT states}
To describe the process of the creation of correlation between two
atoms leading to their entanglement, we need the effective measure
of mixed - states entanglement. As such a measure one usually takes
entanglement of formation $E_{F}(\ro)$ \cite{B}, but in practice it
is not known how to compute this measure for pairs of $d$ - level
systems in the case when $d>2$. A computable measure of entanglement
proposed in \cite{VW} is based on the trace norm of the partial
transposition $\ro^{\mr{PT}}$ of the state $\ro$. From the Peres -
Horodecki criterion of separability \cite{P,3H}, it follows that if
$\ro^{\mr{PT}}$ is not positive, then $\ro$ is not separable and one
defines the \textit{negativity} of the state $\ro$ as
\begin{equation}
N(\ro)=\frac{||\ro^{\mr{PT}}||-1}{2}\label{neg}
\end{equation}
$N(\ro)$ is equal to the absolute value of the sum of negative
eigenvalues of $\ro^{\mr{PT}}$ and is an entanglement monotone, but
it cannot detect bound entangled states \cite{H}. Using the measure
(\ref{neg}), one can check that generalized Dicke states are indeed
not maximally entangled, since
$$
N(\ket{s_{kl}})=N(\ket{a_{kl}})=\frac{1}{2}
$$
\par
In this section we study negativity of the asymptotic states
(\ref{roas}). For such initial states, where only the populations in
antisymmetric Dicke states are non - zero, it is possible to obtain
analytic expression for asymptotic negativity. By a direct
calculation one shows that
\begin{equation}
N(\ras)=\frac{1}{2}\,\left[\sqrt{4
(x^{2}+y^{2})+t^{2}}-t\,\right]\label{diag}
\end{equation}
Notice that  every non - trivial asymptotic state from this class is
entangled. On the other hand, asymptotic negativity for initial
states with non -zero coherences, can only be studied numerically.
\subsection{Pure separable initial states}
We start with pure  states (\ref{basis}). It is obvious that the
initial states $\ket{1_{A}}\otimes\ket{1_{B}}$ and
$\ket{2_{A}}\otimes \ket{2_{B}}$ decay to the ground state
$\ket{g}$. On the other hand, the initial state
$\ket{1_{A}}\otimes\ket{3_{B}}$ (atom A in the excited state and
atom B in the ground state) has the population in the Dicke state
$\ket{a_{13}}$ which is equal to $\frac{1}{2}$, thus for  that state
$$
x=\frac{1}{4},\quad t=\frac{1}{2}\quad\text{and}\quad y=z=w=v=0
$$
and the asymptotic state is entangled with negativity
\begin{equation}
N(\ras)=\frac{\sqrt{2}-1}{4}\label{nas1}
\end{equation}
Similarly, the state $\ket{2_{A}}\otimes\ket{3_{B}}$ has the
population $\frac{1}{2}$ in the state $\ket{a_{23}}$ and also
produces asymptotic state with the same value of entanglement. The
same behaviour can be observed for the initial states $
\ket{3_{A}}\otimes\ket{1_{B}}$ and $\ket{3_{A}}\otimes\ket{2_{B}}$.
\par
When two atoms are initially in different excited states i.e. we
have the states $\ket{1_{A}}\otimes\ket{2_{B}}$ or
$\ket{2_{A}}\otimes\ket{1_{B}}$, then the initial populations in the
states $\ket{a_{13}}$ and $\ket{a_{23}}$ are equal to zero, but the
population in the non - stable state $\ket{a_{12}}$ is non - zero
and equals $\frac{1}{2}$. During the evolution this population is
transformed into the states $\ket{a_{13}}$ and $\ket{a_{23}}$, in
such a way that the asymptotic state $\ras$ satisfies
$$
\bra{a_{12}}\ras\ket{a_{12}}=0
$$
and
$$
\bra{a_{13}}\ras\ket{a_{13}}=\bra{a_{23}}\ras\ket{a_{23}}=\frac{1}{4}
$$
Thus the state $\ras$ is also entangled, but its negativity is less
then (\ref{nas1}) and equals $(\sqrt{6}-2)/8$.
\par
Interesting examples of pure non - diagonal initial state are given
by the following superpositions of states
$\ket{1_{A}}\otimes\ket{2_{B}}$ and $\ket{1_{A}}\otimes\ket{3_{B}}$:
\begin{equation}
\Psi_{\phi}=\cos\phi\,
\ket{1_{A}}\otimes\ket{2_{B}}+\sin\phi\,\ket{1_{A}}\otimes\ket{3_{B}},\quad
\phi \in [0,\pi/2]\label{superposition}
\end{equation}
In that case, the initial populations
$$
\bra{a_{13}}P_{\Psi_{\phi}}\ket{a_{13}}=\frac{1}{2}\sin^{2}\phi,\quad
\bra{a_{23}}P_{\Psi_{\phi}}\ket{a_{23}}=0
$$
and
$$
\bra{a_{12}}P_{\Psi_{\phi}}\ket{a_{12}}=\frac{1}{2}\cos^{2}\phi
$$
are transformed into
\begin{equation*}
\begin{split}
&\bra{a_{13}}\ras\ket{a_{13}}=\frac{1}{2}\sin^{2}\phi+\frac{1}{4}\cos^{2}\phi=
\frac{1}{8}\left(3-\cos 2\phi\right)\\
&\bra{a_{23}}\ras\ket{a_{23}}=\frac{1}{4}\cos^{2}\phi\\
&\bra{a_{12}}\ras\bra{a_{12}}=0
\end{split}
\end{equation*}
In that case, the asymptotic states are parametrized by
$$
x=\frac{1}{16}\,\left(3-\cos 2\phi\,\right),\quad
y=\frac{1}{8}\,\cos^{2}\phi,\quad t=\frac{1}{2},\quad
v=-\frac{1}{8}\,\sin 2\phi
$$
and their negativity can be computed numerically. In FIG.1, we plot
the asymptotic negativity as a function of  $\phi$. This figure
shows that for some values of the parameter $\phi$, the
superposition (\ref{superposition}) can have larger asymptotic
negativity then that achieved by the initial state
$\ket{1_{A}}\otimes\ket{3_{B}}$, which has the maximal value of
negativity produced by the dynamics for pure diagonal initial
states.
\begin{figure}[t]
\centering
{\includegraphics[height=52mm]{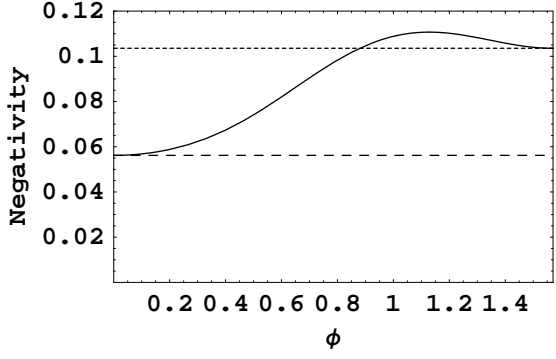}}\caption{Asymptotic
negativity of (\ref{superposition}) as a function of $\phi$. The
values $(\sqrt{6}-2)/8$ and $(\sqrt{2}-1)/4$ are indicated by dashed
and dotted lines respectively. }
\end{figure}
\subsection{Some mixed separable initial states}
\begin{figure}[t]
\centering
{\includegraphics[height=52mm]{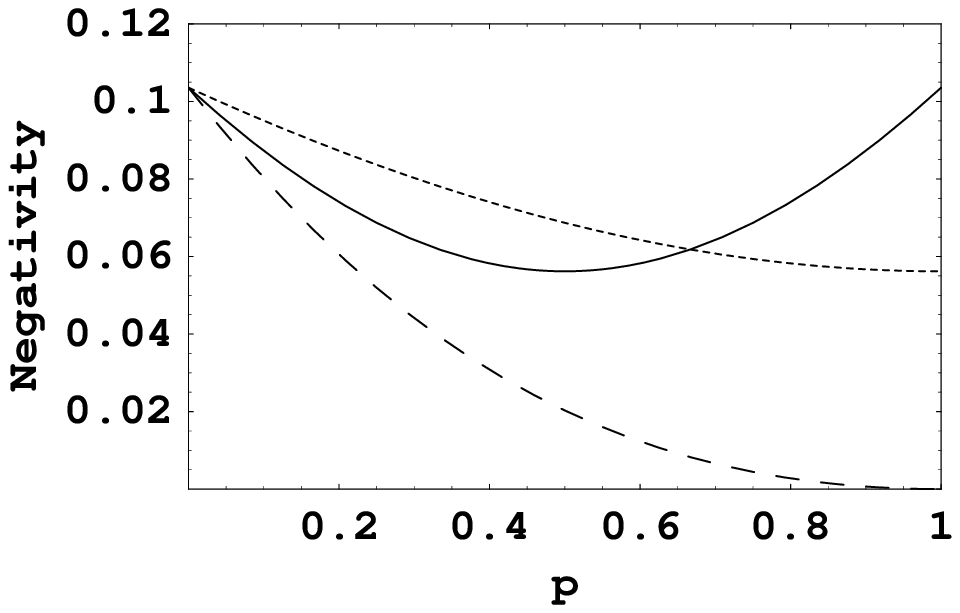}}\caption{Asymptotic
negativity for the mixtures of states
$\ket{1_{A}}\otimes\ket{2_{B}}$ and $\ket{1_{A}}\otimes\ket{3_{B}}$
(dotted line); $\ket{1_{A}}\otimes\ket{1_{B}}$ and
$\ket{1_{A}}\otimes\ket{3_{B}}$ (dashed line);
$\ket{1_{A}}\otimes\ket{3_{B}}$ and $\ket{2_{A}}\otimes \ket{3_{B}}$
(solid line).}
\end{figure}
For the incoherent mixtures of pure states
$\ket{1_{A}}\otimes\ket{2_{B}}$ and $\ket{1_{A}}\otimes\ket{3_{B}}$
i.e. the initial states
\begin{equation}
\ro=p\,\ket{1_{A}}\otimes\ket{2_{B}}\bra{1_{A}}\otimes\bra{2_{B}}+(1-p)\,
\ket{1_{A}}\otimes\ket{3_{B}}\bra{1_{A}}\otimes\bra{3_{B}}\label{2mixed}
\end{equation}
the dynamics also produces entangled asymptotic states. One can
check that their negativity is given by
\begin{equation}
N(\ras)=\frac{1}{\sqrt{32}}\,\sqrt{4-2p+p^{2}}-\frac{1}{4}\label{mixedneg}
\end{equation}
and observe that  (\ref{mixedneg}) as the function of mixing
parameter $p$ is convex, so in contrast to the coherent
superposition, the asymptotic negativity never exceeds (\ref{nas1}).
Analogous properties of negativity can be found for another mixtures
of two pure diagonal states (FIG. 2).
\par
To study the asymptotic entanglement produced for general class of
mixed diagonal states, we consider two parameters to describe
asymptotic states: their negativity and degree of mixture given by
the linear entropy
\begin{equation}
S_{L}(\ro)=\frac{9}{8}\,\tr (\ro-\ro^{2})\label{linentr}
\end{equation}
As we show numerically, the set of all asymptotic states
corresponding to the diagonal initial states is represented on the
entropy - negativity plane by the region $\Lambda_{\mr{as}}$ bounded
by three curves (FIG. 3).
\begin{figure}[b]
\centering {\includegraphics[height=52mm]{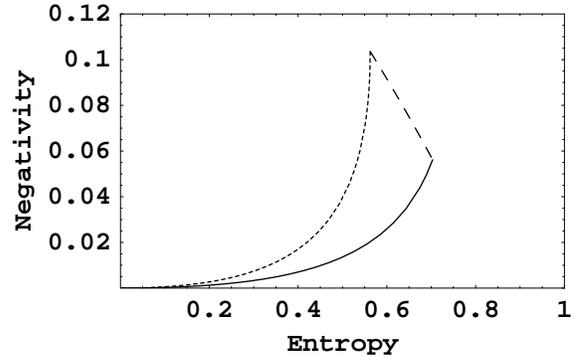}}\caption{The set
$\Lambda_{\mr{as}}$ for the asymptotic states for all diagonal
initial states. The solid line corresponds to the curve (\ref{k1});
the dotted line corresponds to (\ref{k2}) and the dashed line
corresponds to (\ref{k3}).}
\end{figure}
The boundary curves can be found analytically and are given by the
following equations: the solid line on FIG. 3 is described by
\begin{equation}
\begin{split}
&N=\frac{1}{8}\sqrt{2s}-\frac{1}{4}\left(1+\frac{\sqrt{3s-8}-1}{3}\right)\\
&S_{L}=\frac{9}{64}\,(8-s)
\end{split}\label{k1}
\end{equation}
for $s\in [3,8]$, the dotted line is given by
\begin{equation}
\begin{split}
&N=\frac{1}{4}\,\left(\sqrt{2s}-\sqrt{s-1}-1\right)\\
&S_{L}=\frac{9}{16}\,(2-s)
\end{split}\label{k2}
\end{equation}
for $s\in [1,2]$, and finally, the dashed line is described by
equations
\begin{equation}
\begin{split}
&N=\frac{1}{8}\sqrt{2s}-\frac{1}{4}\\
&S_{L}=\frac{9}{64}\,(8-s)
\end{split}\label{k3}
\end{equation}
where $s\in [9/16,45/64]$.
\par
Every asymptotic state corresponding to diagonal initial state is
given by some point from the set $\Lambda_{\mr{as}}$. In particular,
on the curve (\ref{k1}) lie asymptotic states produced from the
mixtures of the state $\ket{1_{A}}\otimes\ket{1_{B}}$ and
$\ket{1_{A}}\otimes\ket{2_{B}}$. Asymptotic states corresponding to
the mixtures of $\ket{1_{A}}\otimes\ket{1_{B}}$ and
$\ket{1_{A}}\otimes\ket{3_{B}}$ lie on the curve (\ref{k2}), whereas
on the curve (\ref{k3}) lie asymptotic states obtained from the
mixtures of states $\ket{1_{A}}\otimes\ket{2_{B}}$ and
$\ket{1_{A}}\otimes\ket{3_{B}}$.
\par
An example of the  state lying inside the set $\Lambda_{\mr{as}}$ is
given by the asymptotic state generated from the initial state
\begin{equation}
\ro_{\infty}=\frac{1}{9}\,\I_{9}\label{maxmix}
\end{equation}
which is maximally mixed state of two qutrits. One can check that
the asymptotic state is given by the formula (\ref{roas}) with
$x=y=1/12$ and $t=2/3$ . The corresponding values of linear entropy
and negativity are equal to $9/16$ and $(3\sqrt{2}-4)/12$,
respectively. As we see, even in that case, the incoherent process
of spontaneous emission produces such correlations which entangle
two atoms and diminish entropy of the system.
\subsection{Initial states with bound entanglement}
Consider now the following initial states
\begin{equation}
\ro_{a}=\frac{1}{8a+1}\begin{pmatrix}a&0&0&0&a&0&0&0&a\\
0&a&0&0&0&0&0&0&0\\
0&0&a&0&0&0&0&0&0\\
0&0&0&a&0&0&0&0&0\\
a&0&0&0&a&0&0&0&a\\
0&0&0&0&0&a&0&0&0\\
0&0&0&0&0&0&\alpha&0&\beta\\
0&0&0&0&0&0&0&a&0\\
a&0&0&0&a&0&\beta&0&\alpha
\end{pmatrix}\label{ppt}
\end{equation}
where
$$
\alpha=\frac{1+a}{2},\quad \beta=\frac{\sqrt{1-a^{2}}}{2}
$$
and $0<a<1$. The states (\ref{ppt}) have positive  partial
transposition, but nevertheless are entangled \cite{H}. Their
entanglement is bound and cannot be distilled \cite{HHH}. It can be
checked by applying for example the realignment criterion of
entanglement \cite{Chen, Rud}. The criterion can be stated as
follows: if the trace norm of realigned state $R(\ro)$ is greater
then $1$, the the state $\ro$ is entangled. We can also introduce
the measure of entanglement based on this criterion. The so called
realignment negativity \cite{BZ} $N_{R}(\ro)$ of the state $\ro$ is
defined by formula
$$
N_{R}(\ro)=||R(\ro)||-1
$$
This measure can detect some bound entangled states, but not all of
them. In the case of states (\ref{ppt}), the values of
$N_{R}(\ro_{a})$ are contained in the interval $(0,\, 0.0035)$ and
the maximal value is attained for $a=\frac{1}{4}$.
\par
Numerical analysis of the evolution of the initial states
(\ref{ppt}) indicates that their realignment negativity very rapidly
goes to zero, but for the later times the states become entangled
with positive values of negativity. Thus the dynamics studied in the
paper has a remarkable property: bound entangled initial states
(\ref{ppt}) evolve into "free" entangled asymptotic states. By a
direct calculation one can show that the asymptotic states  have the
form (\ref{roas}) with parameters
$$
x=\frac{5a+1}{64a+8},\quad y=\frac{3a}{32a+4},\quad
w=-\frac{\sqrt{1-a^{2}}}{32a+4},\quad t=\frac{21a+3}{32a+4}
$$
The values of negativity of those asymptotic states are plotted in
FIG. 4.
\begin{figure}[h]
\centering
{\includegraphics[height=52mm]{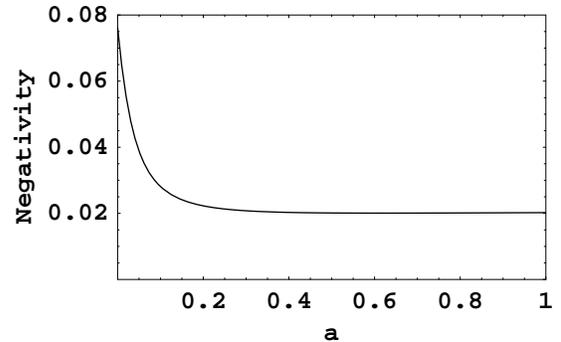}}\caption{Asymptotic
negativity of the initial states (\ref{ppt}) as a function of
parameter $a$.}
\end{figure}
\section{Conclusions}
We have studied entanglement production in the system of two three -
level atoms in the $V$ configuration, coupled to the common vacuum.
In the case of small (compared to the radiation wavelength)
separation between the atoms, the system has nontrivial asymptotic
states which can be entangled even if the initial states were
separable. Particular examples of such separable initial states are
pure states in which the atoms are in different excited states. The
process of the photon exchange between the atoms, produces
correlations that entangle two atoms. It is interesting that when we
superpose  such initial states, we can enlarge the amount the
asymptotic entanglement. We have also characterized the entanglement
of asymptotic states for mixed diagonal initial states. Using the
description of mixed states in terms of degree of mixture and
entanglement, we have found the region on the mixture - entanglement
plane corresponding to such asymptotic states. We have also shown by
considering a specific example that the dynamical evolution of that
system brings bound entangled PPT states into "free" entangled NPPT
asymptotic states.

\end{document}